\documentclass[%
 reprint,
superscriptaddress,
 amsmath,amssymb,
 aps,
]{revtex4-1}

\usepackage{braket}
\usepackage{graphicx}
\usepackage{dcolumn}
\usepackage{bm}
\usepackage{comment}
\usepackage{amsthm}
\theoremstyle{definition}

\newtheorem*{theorem*}{Theorem}

\begin{document}

\preprint{APS/123-QED}

\title{Floquet engineering of topological phases \\ protected by emergent symmetries under resonant drives}

\author{Kaoru Mizuta}
 \email{mizuta.kaoru.65u@st.kyoto-u.ac.jp}
\affiliation{%
 Department of Physics, Kyoto University, Kyoto 606-8502, Japan
}%
\author{Kazuaki Takasan}%
\affiliation{%
 Department of Physics, University of California, Berkeley, California 94720, USA
}%
\author{Norio Kawakami}
\affiliation{%
 Department of Physics, Kyoto University, Kyoto 606-8502, Japan
}%

\date{\today}
             

\begin{abstract}
Floquet engineering is one of the most vigorous fields in periodically driven (Floquet) systems, with which we can control phases of matter usually by high-frequency drives.  In this paper, with Floquet engineering by a combination of high-frequency drives and resonant drives, we propose a way to realize nontrivial topological phases protected by a $\mathbb{Z}_2 \times \mathbb{Z}_2$ symmetry only in the presence of a $\mathbb{Z}_2$ symmetry, using a robust emergent $\mathbb{Z}_2$ symmetry induced by the resonant drives. Moreover, the symmetry protected topological (SPT) phases are switchable between nontrivial and trivial phases only by the direction of a static transverse field, and even perturbations on the resonant drive can be utilized to realize richer SPT phases. We also discuss the real-time dynamics of the model, and find that which topological phases the system lies in can be distinguished by a period doubling of a nonlocal order parameter, as with discrete time crystals.  A realization or a control of nontrivial SPT phases without the required symmetries by resonant drives, proposed in this paper, would shed a new light on the observation of topological phenomena in nonequilibrium setups.

\end{abstract}

\pacs{Valid PACS appear here}
\maketitle


\section{Introduction} 
Periodically driven (Floquet) systems, where the Hamiltonian varies periodically in time, have attracted much interest as one of the most important classes of nonequilibrium systems. For instance, Floquet systems can host unique phases which have no counterparts in equilibrium systems, such as anomalous Floquet topological phases \cite{Kitagawa2010,Rudner2013,Nathan2015,Titum2016,Po2016,Mukherjee2017} and discrete time crystals \cite{Else2016,Keyserlingk2016,Khemani2016,Khemani2017,Yao2017}. One of the most vigorous fields in Floquet systems is \textit{Floquet engineering}, which enables us to control phases of matter with a periodic drive \cite{Oka2009,Kitagawa2011,Lindner2011,Grushin2014,Wang2013,Jotzu2014}. A typical example is a photo-induced topological insulator, which is made topologically nontrivial by a laser light in spite of triviality in the undriven case \cite{Oka2009,Kitagawa2011}. Moreover, it has been revealed that Floquet engineering is applicable also to many-body systems \cite{Kuwahara2016,Mori2016,Abanin2017B,Abanin2017Mat}, though driven many-body systems are in general believed to be thermalized to trivial infinite temperature states \cite{Lazarides2014,DAlessio2014}. This is because, under high-frequency drivings, energy absorption is suppressed and effective static systems are realized in intermediate long time regime based on the high-frequency theories. Thus, Floquet engineering in many-body systems have also attracted much interest \cite{Takasan2017}.

Floquet engineering is achieved by modulating the system with high-frequency drivings and obtaining a desirable effective static Hamiltonian based on the high-frequency expansion \cite{Bukov2015,Eckardt2015,Mikami2016}. Floquet engineering basically aims to control phases by high-frequency drivings, but it turns out that we can add symmetries to a given system with a periodic drive in certain situations \cite{Iadecola2015,Potirniche2017,Agarwal2019}. In particular, recent studies have revealed that, by using resonant drives whose energy scale is comparable to the frequency, we can add a robust symmetry protected by discrete time translation symmetry \cite{Else2017,Mizuta2019}. These studies imply that we can realize nontrivial symmetry protected topological (SPT) phases even when the required symmetry is not respected in nonequilibrium systems. However, the exploration for nontrivial SPT phases realized and controlled by resonant and high-frequency drives with the robust emergent symmetry has not been done well, though it is expected to open up a new way for observation and application of topological phenomena.

In this paper, as a new insight of Floquet engineering, we propose a scheme to realize and control nontrivial SPT phases protected by a robust emergent symmetry under a resonant drive. To be specific, we consider a $\mathbb{Z}_2$ symmetric system and establish a way to realize nontrivial topological phases protected by $\mathbb{Z}_2 \times \mathbb{Z}_2$ symmetry in such systems, which are robust as long as the $\mathbb{Z}_2$ symmetry and time-periodicity are maintained. Moreover, the topological phases can be easily controlled only by the direction of a static field. This unusual realization of SPT phases is achieved by a resonant drive, which brings an additional robust $\mathbb{Z}_2$ symmetry, and a high-frequency drive, which brings topologically nontrivial terms. We also discuss the effect of perturbations on the resonant drive, and find that there appear three distinct nontrivial SPT phases while only the unique nontrivial phase can be realized otherwise. The real-time dynamics of the model is then explored as a tool for distinguishing all the four distinct phases including a trivial phase. We find that all the phases are identified by a period doubling of a nonlocal order parameter, a la discrete time crystals. We further propose possible experimental platforms for performing such a control of SPT phases, especially focusing on cold atoms. This study will shed a new light on realizing or controlling SPT phases in a nonequilibrium setup.

This paper is organized as follows. In Section II, we briefly explain symmetry protected Floquet engineering, with which we can simultaneously control phases of matter and add symmetries to the system with a resonant drive. The following sections III and IV present the main results of our paper. After describing a driven $\mathbb{Z}_2$ symmetric model (Section III), the model is analyzed based on the method provided in the Section II, and we show how nontrivial topological phases protected by $\mathbb{Z}_2 \times \mathbb{Z}_2$ symmetry are realized in this $\mathbb{Z}_2$ symmetric system (Section IV).  In Section V, we consider the effect of perturbations on the resonant drive, and show that it provides us with a broader way of controlling SPT phases. We also discuss the real-time dynamics of the model, which can distinguish all of the SPT phases realized in the model in Section VI. Finally, we provide some promising experimental platforms using the ultracold atoms in optical lattices and summarize our paper in Section VII.

\section{Symmetry protected Floquet engineering}
First of all, we would like to briefly review how a simultaneous control of phases and symmetries can be performed in Floquet systems under resonant drives \cite{Mizuta2019}. We consider Floquet systems with the period $T$, and assume that the Hamiltonian $H(t)$ is composed of time-periodic drives with two local energy scales. Namely, the Hamiltonian $H(t)$ is given by $T$-periodic Hamiltonians $H_0(t)$, whose local energy scale is comparable to the frequency $\omega \equiv 2\pi/T$ (resonant drive), and $V(t)$, whose local energy scale is much smaller than the frequency $\omega$ (high-frequency drive), as follows:
\begin{equation}\label{assumption1}
H(t)=H_0(t)+V(t).
\end{equation}
Then, the resonant drive $H_0(t)$ is assumed to satisfy the following condition for a certain $N \in \mathbb{N}$,
\begin{equation}\label{assumption2}
X^N=1, \quad X \equiv \mathcal{T} \exp \left( -i \int_0^T H_0(t) dt \right).
\end{equation}
This equation represents that the resonant Hamiltonian $H_0(t)$ induces a local $\mathbb{Z}_N$ symmetry operation $X$ on the system every period. Under these assumptions, by performing a perturbative expansion in terms of $\lambda NT$ where $\lambda$ is the local energy scale of $V(t)$, the time evolution operator for the coarse-grained stroboscopic dynamics at $t=mNT \, (m \in \mathbb{N})$ can be described by an effective static Hamiltonian $D_n$,
\begin{equation}\label{result1}
U(NT) \equiv \mathcal{T} \exp \left( -i \int_0^{NT} H(t) dt \right) \sim e^{-iD_n NT}.
\end{equation}
Here, the symbol $\sim$ means that both hand sides are unitarily equivalent approximately, and $D_n$ is the $n$-th order van Vleck effective Hamiltonian given by the following formula,
\begin{eqnarray}
D_n &=& \sum_{i=1}^{n} D_{\mathrm{vV}}^{(n)} \label{D_n}\\
D_\mathrm{vV}^{(1)} &=& V_0, \label{vanVleck1} \\
D_\mathrm{vV}^{(2)} &=& N\sum_{m\neq0}\frac{[V_{-m},V_m]}{2m\omega}, \label{vanVleck2} \\
D_\mathrm{vV}^{(3)} &=& N^2\sum_{m\neq0}\frac{[[V_{-m},V_0],V_m]}{2m^2\omega^2} \nonumber \\
&\quad& +N^2\sum_{m\neq0}\sum_{n\neq0,m} \frac{[[V_{-m},V_{m-n}],V_n]}{3mn\omega^2} \label{vanVleck3}, \\
&\vdots& \nonumber 
\end{eqnarray}
where the $m$-th Fourier component $V_m$ is obtained by
\begin{eqnarray}
V_m &=& \frac{1}{NT} \int^{NT}_0 U_0^\dagger(t)V(t)U_0(t) e^{im\omega t/N} dt, \label{Fourier} \\
U_0(t) &=& \mathcal{T} \mathrm{exp} \left( -i \int^t_0 H_0(t')dt' \right). \label{U_0}
\end{eqnarray}

Most importantly, it is proven that the effective Hamiltonian $D_n$ acquires the new emergent $\mathbb{Z}_N$ symmetry $X$ for any truncation order $n \in \mathbb{N}$, which is described by
\begin{equation}\label{result2}
X D_n X^{-1} = D_n, \qquad \,^\forall n \in \mathbb{N},
\end{equation}
although the original system described by $H(t)$ does not necessarrily respect the symmetry $X$. It is also clarified that the timescale for which the approximation by Eq. (\ref{result1}) works well (called prethermal regime) exponentially grows with  the renormalized frequency $\tilde{\omega} \equiv 1/\lambda NT$ and thus an effective static system described by $D_n$ with the emergent $\mathbb{Z}_N$ symmetry $X$ is realized in this prethermal regime.

Symmetry protected Floquet engineering is performed by a resonant drive and a high-frequency drive based on the above formulation. Then, we can obtain the effective static system described by $D_n$ with the new emergent symmetry $X$. By utilizing the higher order correction terms in Eq. (\ref{D_n}) and the emergent $Z_N$ symmetry which appears in any order term represented by Eq. (\ref{result2}), we can simultaneously control phases of matter and add a robust unitary $\mathbb{Z}_N$ symmetry to the system. This implies that we can control SPT phases even when the original system does not respect the required symmetry. In the following sections in this paper, we consider a $\mathbb{Z}_2$ symmetric model and establish a way to realize nontrivial SPT phases protected by $\mathbb{Z}_2 \times \mathbb{Z}_2$ symmetry with this protocol.

\section{Setup and Model}
In this section, we discuss the setup and the model, with which we can control SPT phases protected by a $\mathbb{Z}_2 \times \mathbb{Z}_2$ symmetry under a $\mathbb{Z}_2$ symmetry. First of all, we would like to clarify which symmeties are focused on throughout this paper. Let us consider a one-dimensional Ising spin chain with a global $\mathbb{Z}_2$ symmetry denoted by
\begin{equation}\label{X_all}
X_\mathrm{all} = \prod_{j=1}^L \sigma_j^x.
\end{equation}
Then, we aim to realize nontrivial topological phases protected by a $\mathbb{Z}_2 \times \mathbb{Z}_2$ symmetry generated by $X_\mathrm{o}$ and $X_\mathrm{e}$, where
\begin{equation}\label{X_odd_even}
X_\mathrm{o} = \prod_{j:\mathrm{odd}} \sigma_j^x, \quad X_\mathrm{e} = \prod_{j:\mathrm{even}} \sigma_j^x.
\end{equation}
The strategy is based on the method provided in Section II, that is, the $\mathbb{Z}_2$ symmetry $X_\mathrm{o}$ is added to the system by a resonant drive. Then, due to the existence of the $\mathbb{Z}_2$ symmetry $X_\mathrm{all}$, the effective static system also respects the other $\mathbb{Z}_2$ symmetry $X_\mathrm{e}=X_\mathrm{all} \cdot X^{-1}_\mathrm{o}$, and thereby becomes a platform for SPT phases under a $\mathbb{Z}_2 \times \mathbb{Z}_2$ symmetry.

\begin{figure}
\begin{center}
    \includegraphics[height=3.75cm, width=8.5cm]{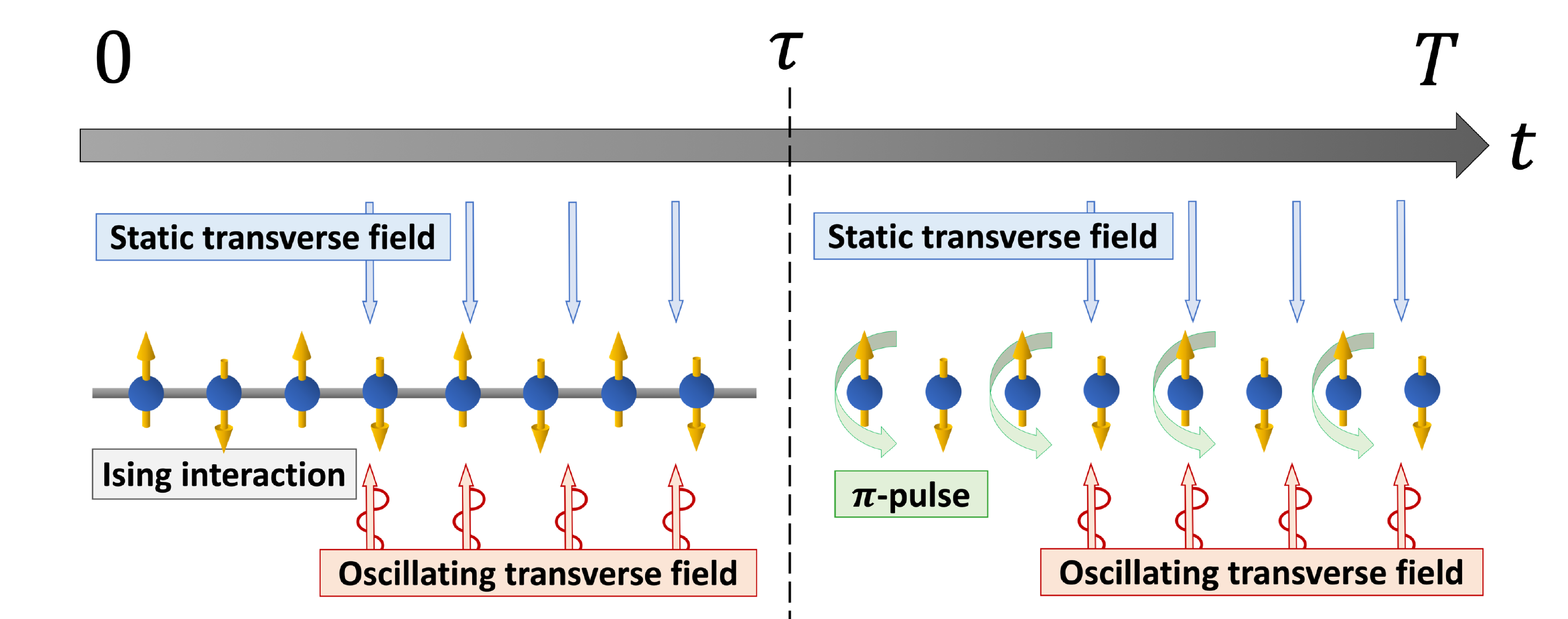}
    \caption{Schematic picture of the driving protocol over one period $T$. The Ising interaction $V_\mathrm{int}(t)$ [Eq. (\ref{V_int})] is uniformly imposed during $0\leq t < \tau$. The resonant drive $H_0(t)$ [Eq. (\ref{H_0_g})], which acts as a $\pi$-pulse, is imposed only on the odd sites during $\tau \leq t < T$. The oscillating transverse field $V_h(t)$ [Eq. (\ref{V_h})] and the static transverse field $V_g$ [Eq. (\ref{V_g})] are both uniformly imposed throughout.}
    \label{Fig_system}
  \end{center}
\end{figure}

Let us describe the model of a one-dimensional Ising spin chain. We assume that the number of sites, $L$, is a multiple of four \cite{NofSite} and impose the open boundary condition. We also assume that the system respects a global $\mathbb{Z}_2$ symmetry $X_\mathrm{all}$.
Then, we consider the periodically driven system as follows,
\begin{equation}\label{Hamiltonian}
H(t) =
\begin{cases}
J \mathop{\sum}\limits_{j=1}^{L-1} \sigma_j^z \sigma_{j+1}^z + (g+h \cos \omega t) \mathop{\sum}\limits_{j=1}^L \sigma_j^x \\ \qquad \qquad \qquad \qquad \qquad \qquad \quad (0 \leq t < \tau)\\
\dfrac{\pi}{2(T-\tau)} \mathop{\sum}\limits_{j:\mathrm{odd}}\sigma_j^x +  (g+h \cos \omega t) \mathop{\sum}\limits_{j=1}^L \sigma_j^x \\
\qquad \qquad \qquad \qquad \qquad \qquad \quad (\tau \leq t < T)
\end{cases}
\end{equation}
Here, to perform the van Vleck expansion [Eq. (\ref{D_n})], the interaction $J$ and the transverse field $g$ and $h$ are assumed to be small compared to the frequency $\omega \equiv 2\pi/T$. Let us rewrite the Hamiltonian Eq. (\ref{Hamiltonian}) so that resonant and high-frequency drives can be distinguished as follows:
\begin{eqnarray}
H(t) &=& H_0(t) + V_\mathrm{int}(t) + V_h(t) + V_g \label{Hamiltonian_r}\\
H_0(t) &=&
\begin{cases}
0 & (0 \leq t < \tau)\\
\dfrac{\pi}{2(T-\tau)} \mathop{\sum}\limits_{j:\mathrm{odd}}\sigma_j^x & (\tau \leq t < T)
\end{cases} \label{H_0_g} \\
V_\mathrm{int}(t) &=&
\begin{cases}
J \mathop{\sum}\limits_{j=1}^{L-1} \sigma_j^z \sigma_{j+1}^z & (0 \leq t < \tau) \\
0 & (\tau \leq t < T)
\end{cases} \label{V_int} \\
V_h(t) &=& h \cos \omega t \sum_{j=1}^L \sigma_j^x \label{V_h}\\
V_g &=& g \sum_{j=1}^N \sigma_j^x. \label{V_g}
\end{eqnarray}
This model is schematically illustrated in Fig. \ref{Fig_system}. Then, the high-frequency drive $V(t)$ in Eq. (\ref{assumption1}) is given by $V_\mathrm{int}(t)+V_h(t)+V_g$. The resonant drive $H_0(t)$ induces the $\mathbb{Z}_o$ symmetry $X_\mathrm{o}$, that is,
\begin{equation}
\mathcal{T} \exp \left( -i \int_0^T H_0(t) dt \right) = X_\mathrm{o}.
\end{equation}
Thus, the method provided in Section II is applicable with $N=2$.

It should be noted that the transverse fields $H_0(t)$, $V_h(t)$, and $V_g$ respect the $\mathbb{Z}_2 \times \mathbb{Z}_2$ symmetry represented by $X_o$ and $X_e$, but they are topologically trivial since they are onsite operators. The Ising interaction term $V_\mathrm{int}(t)$ does not even respect the $\mathbb{Z}_2 \times \mathbb{Z}_2$ symmetry, since it anticommutes with $X_o$ and $X_e$. Despite their simple forms, they can become sources for realizing nontrivial SPT phases under the $\mathbb{Z}_2 \times \mathbb{Z}_2$ symmetry as we discuss in the next section.

\section{Analysis: Control of SPT phases}
In this section, we would like to discuss how nontrivial SPT phases under the $\mathbb{Z}_2 \times \mathbb{Z}_2$ symmetry are realized or controlled in the system which only respects $\mathbb{Z}_2$ symmetry, described by Eq. (\ref{Hamiltonian}). To this end, we calculate the effective Hamiltonian based on the method in Section II, and explore its topological properties analytically. We also perform a numerical calculation which does not rely on the van Vleck expansion to examine whether or not topological nature obtained by the analytical calculation appears without the perturbative approximation.

We first clarify how topological phases are defined throughout this paper. In Floquet systems, there are several ways to define topological phases \cite{Nakagawa2019}. One of the definitions is based on an effective Hamiltonian defined by $i \log U(T) /T$ \cite{Oka2009,Lindner2011}, where $U(t)$ represents the time evolution operator of the system. The others are defined by a time evolution operator $U(t)$ over one period $T$ \cite{Rudner2013,Nathan2015} or a Floquet operator $U(T)$ \cite{Kitagawa2010,Higashikawa2018}. Our definition is none of them, though it is close to the first one. Here, we focus only on the coarse-grained stroboscopic dynamics $t=2mT$, and define topological phases based on the effective Hamiltonian given by
\begin{equation}\label{H_eff}
H_\mathrm{eff} \equiv \frac{i}{2T} \log U(2T),
\end{equation}
where we choose a branch cut so that all the quasienergies $\{ \varepsilon \}$ (the eigenvalues of $H_\mathrm{eff}$) can be included in $[-\pi/2T,\pi/2T)$. Then, the classification of topological phases based on $H_\mathrm{eff}$ is the same as the one in static systems. Since $H_\mathrm{eff}$ can be unitarily equivalent to the van Vleck effective Hamiltonian $D_n$ approximately from Eq. (\ref{result1}), we can analyze topological phases with the van Vleck effective Hamiltonian $D_n$ instead of the effective Hamiltonian $H_\mathrm{eff}$. We can also define the ground state by the eigenstate of $H_\mathrm{eff}$ with the lowest quasienergy in $[-\pi/2T,\pi/2T)$.

Let us consider the van Vleck effective Hamiltonian $D_n$. To simplify the problem, we assume that there is an energy-scale separation among the high-frequency drivings $V_\mathrm{int}(t)$, $V_h(t)$ and $V_g$. That is, let us suppose that the local energy scale of the former two terms, denoted by $\lambda \equiv \max (J,h)$, is larger than that of the last one, which is described by 
\begin{equation}\label{g_small}
g/\lambda = O \left( \left( \frac{\lambda}{\omega} \right)^2 \right).
\end{equation}
As discussed later, the system hosts a topologically trivial phase when $g$ becomes larger, and hence it is enough to consider within this regime to obtain nontrivial SPT phases.

We perform the approximation by truncating the van Vleck effective Hamiltonian at the third order. With neglecting the third order term originating from $V_g$ due to Eq. (\ref{g_small}), we obtain the effective Hamiltonian $D_3$ as follows,
\begin{eqnarray}
D_3 &=& g \sum_j \sigma_j^x + \gamma \sum_j (\sigma_j^x + \sigma_{j-1}^z \sigma_j^x \sigma_{j+1}^z) \nonumber \\
&\quad& + \,\, (\mathrm{local\,\, terms\,\, on\,\, the\,\, edge}), \label{D_3} \\
\gamma &=&  \frac{4J^2h}{3\pi\omega^2} \{ 2\sin \omega \tau - \omega\tau (1+\cos \omega \tau) \}.
\end{eqnarray}
The detailed calculation for obtaining this effective Hamiltonian is provided in Appendix. Here, $\gamma$ is a real number which depends on the parameters $J$, $h$, and $\omega\tau$. The sign of $\gamma$ is determined by $\omega\tau$, as $\mathrm{sgn}(g) = \mathrm{sgn}(\pi-\omega\tau)$ for $0\leq \omega \tau \leq 2\pi$. As discussed in Section II, though the original Hamiltonian respects only the $\mathbb{Z}_2$ symmetry $X_\mathrm{all}$, the effective Hamiltonian $D_3$ possesses the emergent $\mathbb{Z}_2 \times \mathbb{Z}_2$ symmetry represented by $X_\mathrm{o}$ and $X_\mathrm{e}$. This emergent symmetry is easily checked by the fact
\begin{equation}\label{commutation}
[\sigma_{j-1}^z \sigma_j^x \sigma_{j+1}^z, X_\mathrm{o}] =  [\sigma_{j-1}^z \sigma_j^x \sigma_{j+1}^z, X_\mathrm{e}] = 0, \quad \,^\forall j.
\end{equation}
In the discussions below, we neglect the local term which only acts on the edge $j=1$ or $j=L$ in Eq. (\ref{D_3}), because such a local term becomes irrelevant when we consider the topological nature of the system with a large number of sites $L$.

\begin{figure*}
\hspace{-1cm}
\begin{center}
    \includegraphics[height=5cm, width=18cm]{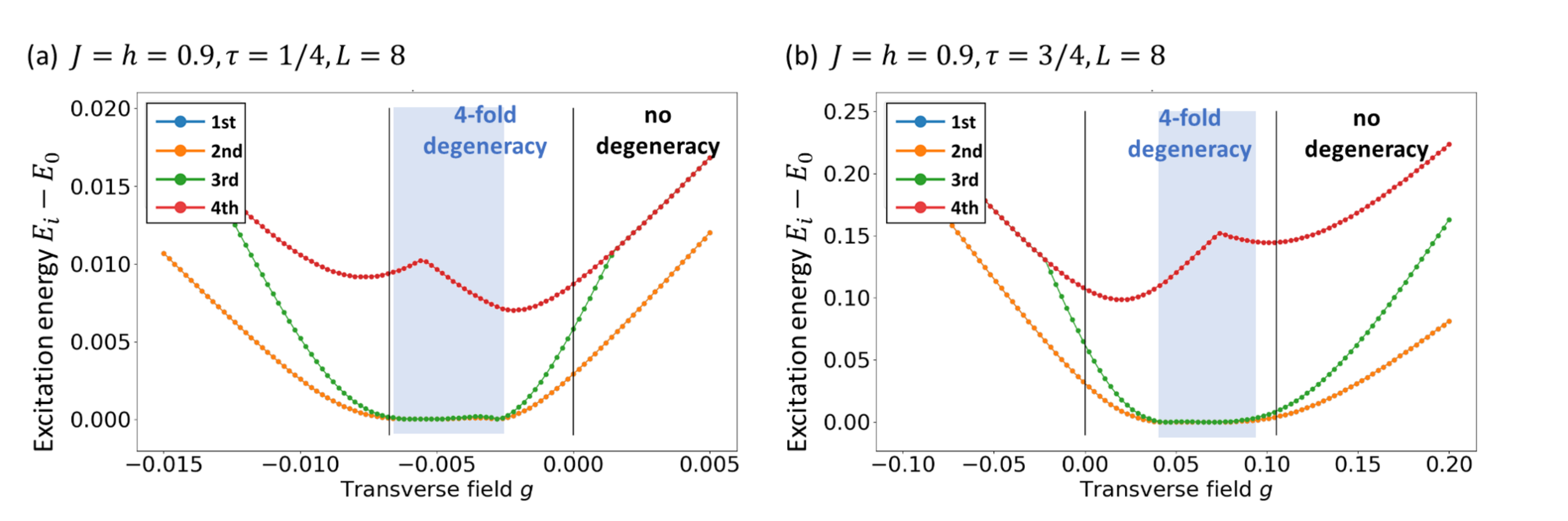}
    \caption{Excitation energies from the ground state to low energy excited states of $H_\mathrm{eff}$ for different choice of switching time (a) $\tau=1/4$ and (b) $\tau=3/4$. In the blue regions, a four-fold degeneracy appears in the ground states, which is a signature of nontrivial SPT phases. The black lines represent the analytically-obtained topological phase transition points $g=0$ and $g=-2\gamma$.  In the parameter regions shown in both graphs (a) and (b), the first excited state (blue dotted line) and the second excited state (orange dotted line) are degenerate.} 
    \label{Fig_g}
  \end{center}
\end{figure*}

Let us discuss how the effective Hamiltonian $D_3$ in Eq. (\ref{D_3}) can be topologically classified and as a result how we can control SPT phases. To this end, we consider a $\mathbb{Z}_2 \times \mathbb{Z}_2$ symmetric Hamiltonian $H$, given by
\begin{equation}\label{Z2_Z2model}
H= a\sum_{j=1}^N \sigma_j^x + b \sum_{j=2}^{N-1} \sigma_{j-1}^z \sigma_j^x \sigma_{j+1}^z
\end{equation}
under the open boundary condition \cite{Iadecola2015}. In fact, this model can be exactly solved by Jordan-Wigner transformation
\begin{equation}\label{JordanWigner}
\alpha_j \equiv \left( \prod_{k<j} \sigma_k^x \right) \sigma_j^z, \quad \beta_j \equiv - \left( \prod_{k<j} \sigma_k^x \right) \sigma_j^y,
\end{equation}
where $\{ \alpha_j \}$ and $\{ \beta_j \}$ are independent Majorana fermion operators which satisfy
\begin{equation}\label{Majorana}
\{ \alpha_i, \alpha_j \} = \{ \beta_i, \beta_j \} = 2 \delta_{ij}, \quad \{ \alpha_i, \beta_j \} =0.
\end{equation}
Then, the static system decribed by $H$ can be mapped to a fermion system, which results in
\begin{equation}\label{KitaevChain}
H = -i a \sum_j \alpha_j \beta_j -i b \sum_j \beta_{j-1} \alpha_{j+1}.
\end{equation}
This model is nothing but two decoupled, independent, and equivalent Kitaev chains \cite{Iadecola2015}---one is composed of the odd sites $j=2m-1$, while the other is composed of the even sites $j=2m$. Due to the equivalence of the two Kitaev chains, there are two possible phases. One is a topologically nontrivial phase when $|a| < |b|$, in which both of the chains become nontrivial. The other phase is a trivial phase under $|a| > |b|$, in which both chains are trivial. The topological phase transition occurs at $|a|=|b|$.

Let us return to the discussion on our model described by Eq. (\ref{D_3}). Note that the second summation term in Eq. (\ref{D_3}) represents a $\mathbb{Z}_2 \times \mathbb{Z}_2$ symmetric Hamiltonian just at the topological phase transition point. Thus, the first term in Eq. (\ref{D_3}) determines topological phases, and assuming $h>0$, the following relation is obtained depending on the sign of $\gamma$.

\begin{flushleft}
When $0<\tau<T/2$, or equivalently $\gamma>0$,
\begin{eqnarray*}
g<-2\gamma, \, 0<g &:& \text{topologically trivial phase}, \\
g=-2\gamma, \, 0 &:& \text{topological phase transition point}, \\
-2\gamma<g<0 &:& \text{topologically nontrivial phase}.
\end{eqnarray*}
\end{flushleft}
\begin{flushleft}
When $T/2<\tau<T$, or equivalently $\gamma<0$,
\begin{eqnarray*}
g<0, \, -2\gamma<g &:& \text{topologically trivial phase}, \\
g=0, \, -2\gamma &:& \text{topological phase transition point}, \\
0<g<-2\gamma &:& \text{topologically nontrivial phase}.
\end{eqnarray*}
\end{flushleft}
Therefore, by choosing a value between $-2\gamma$ and $0$ for the strength of the static transverse field $g$, a nontrivial SPT phase under the $\mathbb{Z}_2 \times \mathbb{Z}_2$ symmetry is realizable, although the original system only respects the $\mathbb{Z}_2$ symmetry (and discrete time translation symmetry), and is driven by $\mathbb{Z}_2 \times \mathbb{Z}_2$ symmetric trivial fields and a $\mathbb{Z}_2 \times \mathbb{Z}_2$ asymmetric interaction. In a topologically nontrivial phase, since each of the two Kitaev chains of the mapped fermionic system has one gapless excitation under the open boundary condition, the ground state of the system has a topologically protected four-fold degeneracy. 

The above condition for the topological phase transition also implies an easy control of SPT phases. Let us consider the small field limit $g \simeq 0$. Then, topological phases only depend on the direction of the static transverse field, and hence, by changing its direction, we can easily switch the system between a trivial phase and a nontrivial SPT phase.

We also numerically examine whether or not the topological nature can appear in the system without resorting to the high-frequency expansion. We calculate the spectrum of the approximation-free effective Hamiltonian $H_\mathrm{eff}$, given by Eq. (\ref{H_eff}), using the exact diagonalization. Figure \ref{Fig_g} shows excitation energies from the ground state to low energy excited states under the Hamiltonian $H_\mathrm{eff}$ for $J=h=0.9, T=1, L=8$. For $g$ in the blue regions, the excitation energies from the ground state to the first, second and third excited states are zero, implying that there is a four-fold degeneracy in the ground state of $H_\mathrm{eff}$. Thus, though the topological phase transition points deviate from those of the analytical calculation, we can conclude that nontrivial topological phase transitions appear by tuning the transverse field $g$.  

The deviation of the phase transition points can be attributed to the finite size effect and the higher order terms which are naturally included in $H_\mathrm{eff}$. Here, we discuss the influence of the higher-order terms of the van Vleck expansion. The way of Floquet engineering with a resonat drive, provided in the Section II, ensures that the emergent $\mathbb{Z}_2$ symmetry $X_\mathrm{e}$ appears in any order term. Since the original Hamiltonian $H(t)$ respects the $\mathbb{Z}_2$ symmetry $X_\mathrm{all}$, the effective Hamiltonian including higher order terms always respects the $\mathbb{Z}_2 \times \mathbb{Z}_2$ symmetry represented by $X_\mathrm{o}$ and $X_\mathrm{e}$. Thus, the higher order terms merely shift the phase transition points and do not break the required $\mathbb{Z}_2 \times \mathbb{Z}_2$ symmetry for realizing SPT phases. This seems to be why a clear four-fold degeneracy is observed without resorting to the approximation in Fig. \ref{Fig_g}. In a similar way, we can discuss the robustness against perturbations. As long as the perturbations respect the $\mathbb{Z}_2$ symmetry $X_\mathrm{all}$, their effect appears on the the van Vleck effective Hamiltonian $D_n$ as $\mathbb{Z}_2 \times \mathbb{Z}_2$ symmetric terms. Thus, properties of nontrivial SPT phases can appear if such $\mathbb{Z}_2$ symmetric perturbations are small enough. What should be done to observe nontrivial SPT phases protected by the $\mathbb{Z}_2 \times \mathbb{Z}_2$ symmetry is only to maintain the $\mathbb{Z}_2$ symmetry and the discrete time translation symmetry.


\section{Control of phases via flip errors}
Symmetry protected Floquet engineering, described in Section II, seems to require a fine-tuning of the resonant drive $H_0(t)$ due to the condition given in Eq. (\ref{assumption2}). In this section, we evaluate effects of the deviation of the resonant drive on topological phases of our model. As a result, we uncover that tuning the deviation of the resonant drive enables us to realize more various phases than that solely by the static transverse field $g$ in the previous section.

In the previous section, a $\pi$-spin flip on the odd sites is considered as a resonant drive. Here, we assume that the angle of rotation slightly deviates from $\pi$ in the perturbed case. Then, the Hamiltonian in the perturbed case is described as follows,
\begin{eqnarray}
H(t) &=& \tilde{H}_0(t) + V_\mathrm{int}(t)+V_h(t)+V_g, \label{Hamiltonian_epsilon} \\
\tilde{H}_0(t) &=& 
\begin{cases}
0 & (0 \leq t < \tau)\\
\dfrac{\pi(1+\varepsilon)}{2(T-\tau)} \mathop{\sum}\limits_{j:\mathrm{odd}} \sigma_j^x & (\tau \leq t < T),
\end{cases} \label{tilde_Hamiltonian}
\end{eqnarray}
where $\varepsilon$ represents a degree of the deviation from the $\pi$-spin flip. Such an error on the rotation angle is widely considered in the context of discrete time crystals and their related phenomena, and it often breaks time-crystalline orders when it increases to some extent \cite{Russomanno2017DTC,Zhang2017,Choi2017,Ho2017,Huang2018,Fan2019,Chew2019}. Moreover, for quantum systems such as trapped ions, the parameter $\varepsilon$ is tunable.

\begin{figure}
\begin{center}
    \includegraphics[height=3.75cm, width=8.5cm]{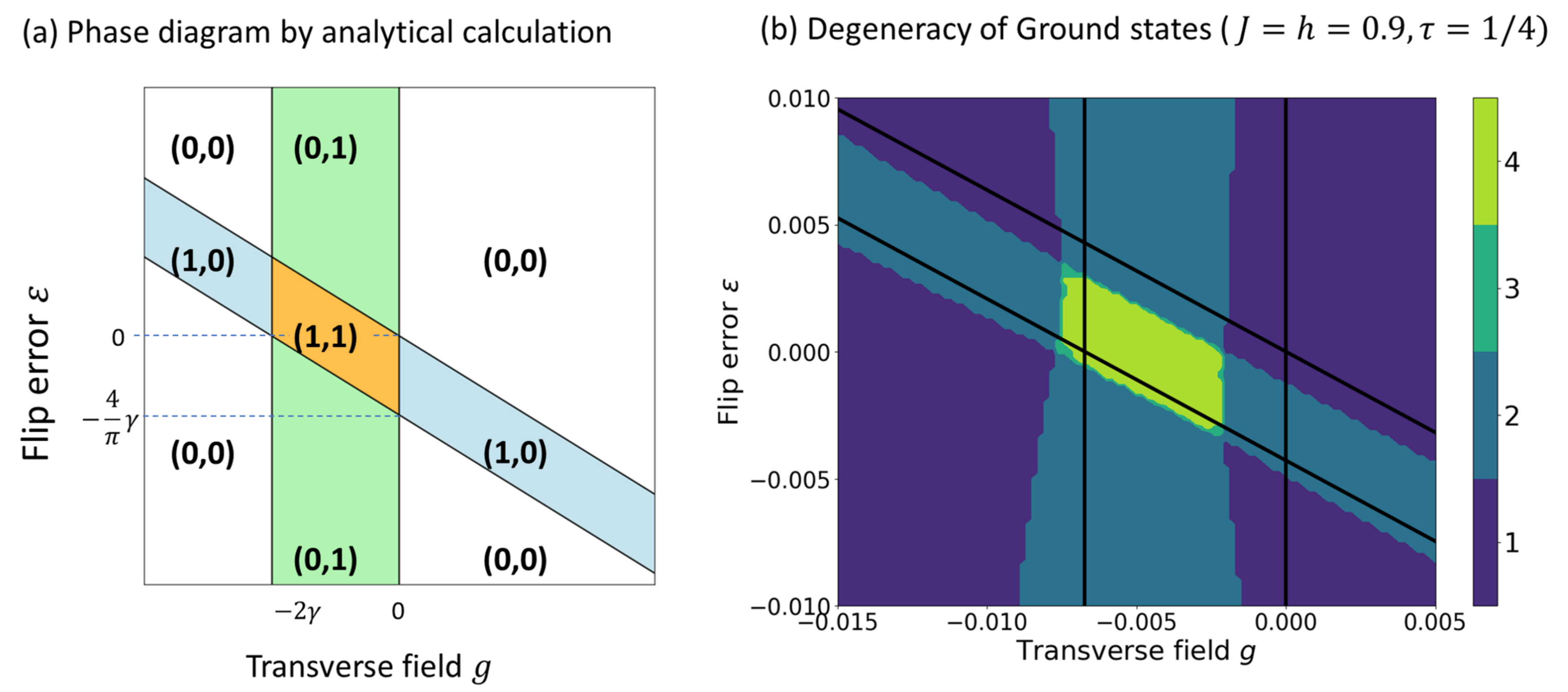}
    \caption{ (a) Phase diagram of the effective static model Eq. (\ref{D_3_epsilon}) based on the van Vleck expansion $D_3$. The indices $(Z_\mathrm{o},Z_\mathrm{e})\in \mathbb{Z}_2 \times \mathbb{Z}_2$ represent the $\mathbb{Z}_2$ topological numbers of the Kitaev chains composed of the odd and even sites, respectively. (b) The degeneracy of the ground states, calculated in terms of the effective Hamiltonian $H_\mathrm{eff}$. Here, we define the degeneracy for the $n$-th excited state with the eigenenergy $E_n$ by satisfying $|E_n-E_0|/|E_0| < 0.01$. The black lines are the boundaries of SPT phases obtained by the analytical calculation of $D_3$. This result well reflects the phase diagram based on the van Vleck expansion in (a).}
    \label{Fig_phase}
  \end{center}
\end{figure}

Let us analyze the perturbed model given by Eqs. (\ref{Hamiltonian_epsilon}) and (\ref{tilde_Hamiltonian}). Since this perturbed resonant drive does not satisfy the condition Eq. (\ref{assumption2}), the method in Section II seems to be inapplicable. However, we can rewrite the Hamiltonian by
\begin{eqnarray}
H(t) &=& H_0(t) + V_\varepsilon(t) + V_\mathrm{int}(t)+V_h(t)+V_g, \label{Hamiltonian_epsilon_r}\\
V_\varepsilon(t) &=& 
\begin{cases}
0 & (0 \leq t < \tau) \\
\dfrac{\pi\varepsilon}{2(T-\tau)} \mathop{\sum}\limits_{j:\mathrm{odd}} \sigma_j^x & (\tau \leq t < T),
\end{cases}\label{V_epsilon}
\end{eqnarray}
and by regarding the high-frequency driving $V(t)$ as $V_\varepsilon(t) + V_\mathrm{int}(t)+V_h(t)+V_g$, the method becomes applicable. Then, we again assume
\begin{equation} \label{epsilon_small}
\varepsilon / \lambda T = O \left( \left( \frac{\lambda}{\omega} \right)^2 \right),
\end{equation}
and neglect the third order term originating from $V_\varepsilon(t)$. This results in the following effective Hamiltonian up to the third order:
\begin{eqnarray}
D_3 &=& a_\mathrm{e} \sum_{j:\mathrm{even}} \sigma_j^x + b_\mathrm{o} \sum_{j:\mathrm{odd}} \sigma_{j-1}^z \sigma_j^x \sigma_{j+1}^z \nonumber \\
&\quad& + a_\mathrm{o} \sum_{j:\mathrm{odd}} \sigma_j^x + b_\mathrm{e} \sum_{j:\mathrm{even}} \sigma_{j-1}^z \sigma_j^x \sigma_{j+1}^z \nonumber \\
&\quad& \quad + \,\, (\mathrm{local\,\, terms\,\, on\,\, the\,\, edge}) \label{D_3_epsilon}
\end{eqnarray}
with $a_\mathrm{o}=g+\pi \varepsilon /2T +\gamma$, $a_\mathrm{e}=g+\gamma$, and $b_\mathrm{o}=b_\mathrm{e}=\gamma$. By neglecting the local term on the edge and performing Jordan Wigner transformation, described by Eq. (\ref{JordanWigner}), this effective Hamiltonian is mapped to the following Majorana-fermionic system,
\begin{eqnarray}
D_3 &=& \left( -ia_\mathrm{o} \sum_m \alpha_{2m-1} \beta_{2m-1} -i b_\mathrm{e} \sum_m \beta_{2m-1} \alpha_{2m+1} \right) \nonumber \\
&\quad& + \left( -ia_\mathrm{e} \sum_m \alpha_{2m} \beta_{2m} -i b_\mathrm{o} \sum_m \beta_{2m} \alpha_{2m+2} \right). \label{KitaevChain_epsilon}
\end{eqnarray}
What is important here is that the effective static system corresponds to two decoupled Kitaev chains but they are not equivalent, and the resulting four distinct phases are characterized by $\mathrm{sgn}(|a_\mathrm{o}|-|b_\mathrm{e}|)$ and $\mathrm{sgn}(|a_\mathrm{e}|-|b_\mathrm{o}|)$.  Therefore, by tuning the transverse field $g$ and the flip error $\varepsilon$, four different phases---a phase where both chains are nontrivial, two phases where one of the two is nontrivial, and a phase where both chains are trivial, are realizable. We show the phase diagram of the system as functions of the transverse field $g$ and the flip error $\varepsilon$ in Fig. \ref{Fig_phase}. The case where both the static transverse field and the flip error are absent, represented by $g=\varepsilon=0$, is just at the boundary of all the four phases. Thus, by slightly changing $(g,\varepsilon)$ from $(0,0)$ in a certain direction, we can realize any of the four phases.

\begin{figure*}
\hspace{-1cm}
\begin{center}
    \includegraphics[height=5cm, width=18cm]{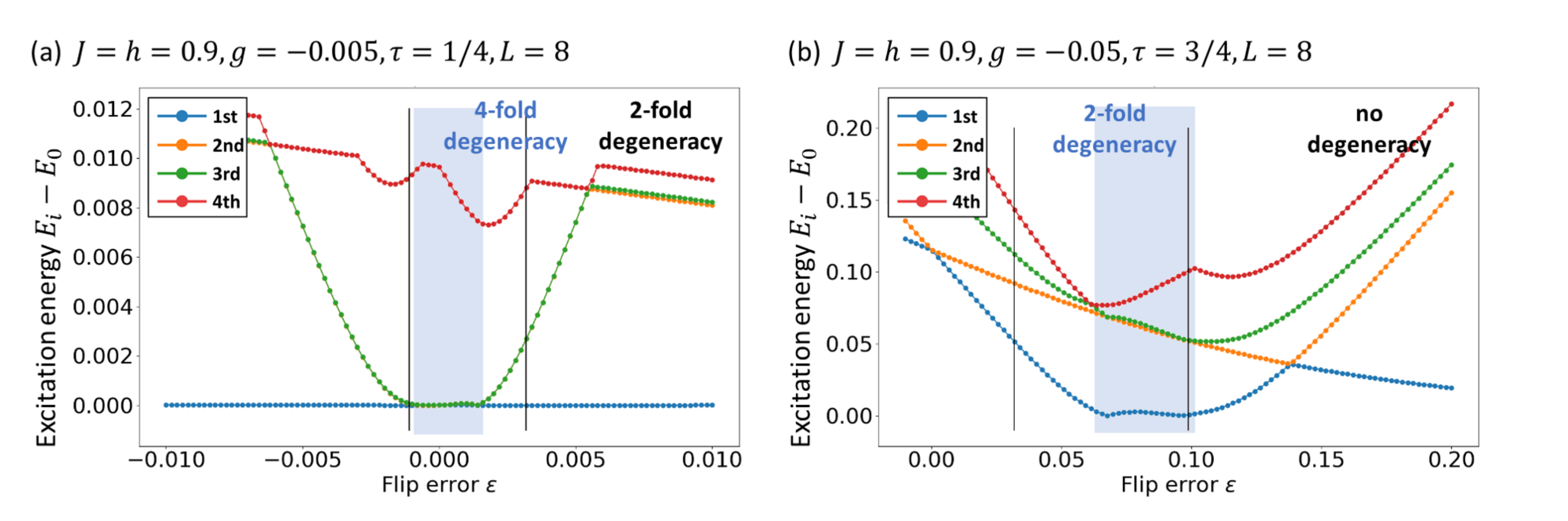}
    \caption{Excitation energies from the ground state to low energy states for (a) $\tau=1/4, g=-0.005$ and (b) $\tau=3/4, g=-0.05$, when the flip error $\varepsilon$ is introduced. In the case of (a), a four-fold degeneracy and a two-fold degeneracy are observed. The second excited state (orange dotted line) and the third excited state (green dotted line) are degenerate in most of the region shown in (a). On the other hand, two-fold-degenerate ground states and non-degenerate ground states are observed in (b).}
    \label{Fig_g_epsilon}
  \end{center}
\end{figure*}

To show the validity of the above discussion which relies on the van Vleck expansion, we also examine the topological phases with numerically calculating the effective Hamiltonian $H_\mathrm{eff}$ defined by Eq. (\ref{H_eff}). Figure \ref{Fig_g_epsilon}  shows the $\varepsilon$-dependence of excitation energies from the ground state to low energy excited states for different choice of switching time (a) $g=-0.005$, $\tau=1/4$ and (b) $g=-0.05$, $\tau=3/4$ under the renormalization $T=1$. In Fig. \ref{Fig_g_epsilon}, there appears a four-fold degeneracy in the ground states within the blue region, which means that both of the Kitaev chains in the mapped system become nontrivial. On the other hand, out of the blue region, a two-fold degeneracy appears in the ground state. From the analytical calculation based on $D_3$ (See Fig. \ref{Fig_phase} (a)), it turns out that the Kitaev chain composed of the even sites in the mapped system is nontrivial while the other is trivial in this region. In Fig. \ref{Fig_g_epsilon} (b), the ground state with a 2-fold degeneracy or without degeneracies appears according to the parameters. Figure \ref{Fig_phase} (b) shows the degeneracy of the ground state for each parameter $g$ and $\varepsilon$, calculated based on $H_\mathrm{eff}$ defined by Eq. (\ref{H_eff}). The yellow and blue regions represent that there are four-fold and two-fold degeneracies in the ground states respectively. In the purple regions, the ground state is not degenerate.  This result well reproduces the phase diagram based on the van Vleck expansion (See Fig. \ref{Fig_g_epsilon} (b)).

\begin{figure*}
\hspace{-1cm}
\begin{center}
    \includegraphics[height=9.5cm, width=18cm]{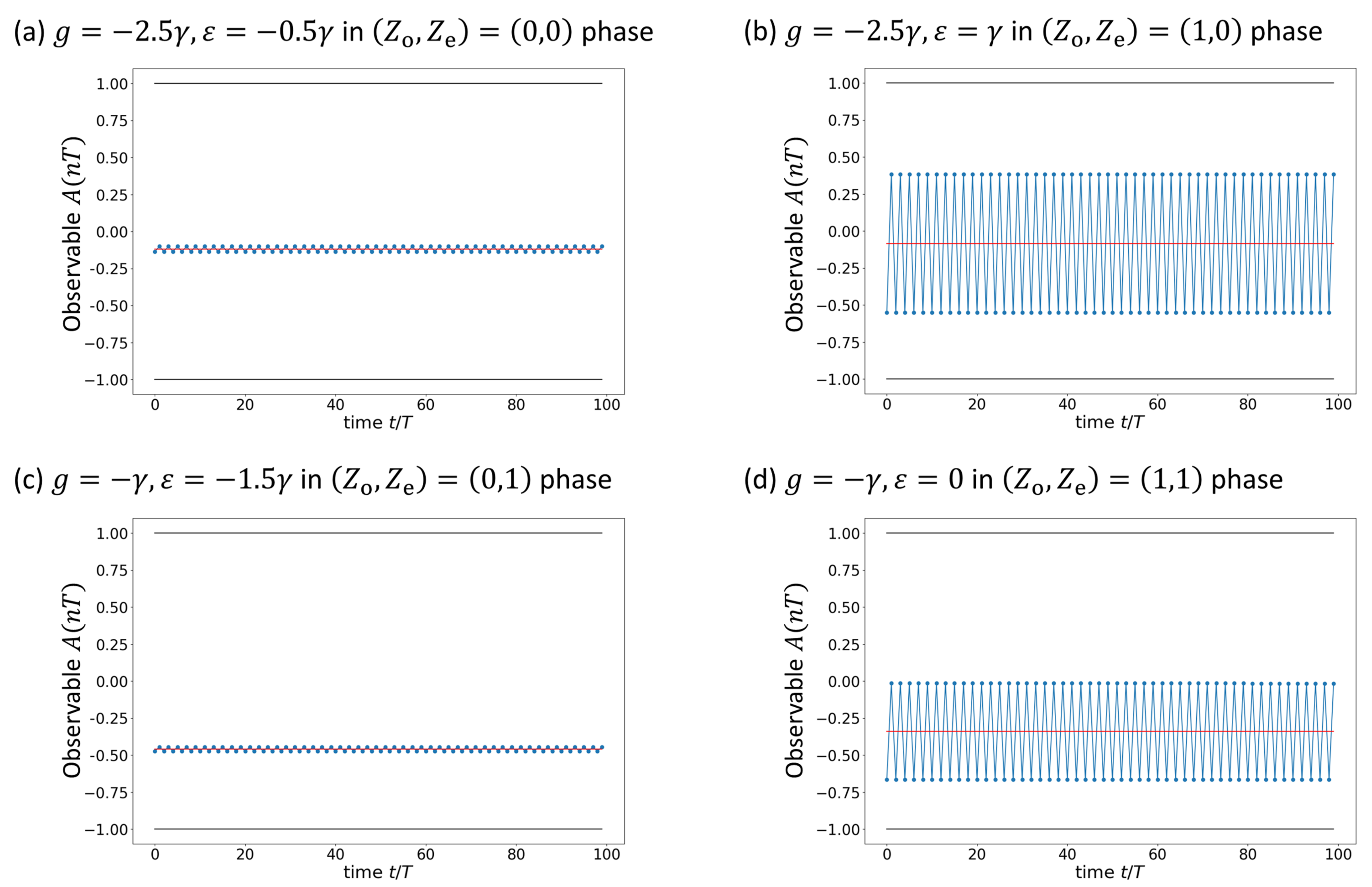}
    \caption{Numerical results of the stroboscopic dynamics of the order parameter $A(nT)$ for different parameters which lie in the four distinct phases. In each graph, we choose $J=0.5, h=-0.5, \tau=1/4$, and $L=8$. The red solid lines represent the mean value of $A(nT)$.}
    \label{Fig_dynamics}
  \end{center}
\end{figure*}

\section{Dynamical signatures of the phases}

In the previous sections, it is shown that, with tuning the static transverse field and the flip error, the system can host four kinds of topological phases protected by $\mathbb{Z}_2 \times \mathbb{Z}_2$ symmetry. In this section, we discuss how the characteristics of the four distinct phases appears in the real-time dynamics.

First of all, we discuss the model based on the third-order van Vleck effective Hamiltonian $D_3$, which is given by Eq. (\ref{D_3}). Using Jordan-Wigner transformation, described by Eq. (\ref{JordanWigner}), this effective model is mapped to two decoupled Kitaev chains. Since one Kitaev chain can be mapped to one transverse Ising chain by Jordan-Wigner transformation, we can map the effective model to two decoupled transverse Ising chains. To be precise, let us define new Pauli operators $\{ \tau_m^{\alpha=x,y,z}\}$ and $\{ \tilde{\tau}_m^{\alpha=x,y,z}\}$ by
\begin{eqnarray}
\tau_m^x &=& \sigma_{2m-1}^x, \quad \tau_m^y = i \tau_m^z \tau_m^x, \\
\tau_m^z &=& 
 \left( \prod_{n < m} \sigma_{2n}^x \right) \sigma_{2m-1}^z \\
\tilde{\tau}_m^x &=& \sigma_{2m}^x, \quad \tilde{\tau}_m^y = i \tilde{\tau}_m^z \tilde{\tau}_m^x, \\
\tilde{\tau}_m^z &=& 
 \left( \prod_{n \leq m} \sigma_{2n-1}^x \right) \sigma_{2m}^z,
\end{eqnarray}
and then the effective model $D_3$ becomes equivalent to two decoupled transverse Ising chains as follows:
\begin{equation}
D_3 = \sum_m \left( a_\mathrm{o} \tau_m^x +b_\mathrm{e} \tau_m^z \tau_{m+1}^z \right)
+  \sum_m \left( a_\mathrm{e} \tilde{\tau}_m^x +b_\mathrm{o} \tilde{\tau}_m^z \tilde{\tau}_{m+1}^z \right)
\end{equation}
with neglecting the terms which were originally the local terms on the edge.

Next, consider a small perturbation which breaks the $\mathbb{Z}_2$ symmetry $X_\mathrm{all}$ and the discrete time translation symmetry ($T$-periodicity), and here, we choose the following $2T$-periodic perturbation
\begin{equation}
H_\mathrm{per}(t) = \begin{cases}
h_z \mathop{\sum}\limits_{m}^{L/2} ( \tau_m^z + \tilde{\tau}_m^z ) & (0 \leq t < \tau) \\
0 & (\tau \leq t < T) \\
h_z \mathop{\sum}\limits_{m}^{L/2} ( \tau_m^z - \tilde{\tau}_m^z ) & (T \leq t < T+\tau) \\
0 & (T+\tau \leq t < 2T).
\end{cases}
\end{equation}
The effective Hamilotnian $\tilde{H}_\mathrm{eff}$ of the perturbed system is given by
\begin{eqnarray}
\tilde{H}_\mathrm{eff} &\equiv& \frac{i}{2T} \log \tilde{U}(2T), \\
\tilde{U}(2T) &=& \mathcal{T} \exp \left\{ -i \int_0^{2T} (H(t)+H_\mathrm{per}(t))dt \right\}.
\end{eqnarray}
We can approximately obtain the effective Hamiltonian by the method in the Section II also in the perturbed case by regarding the system as a $2T$-periodic system and taking $H_0(t)$ as a resonant term which satisfies $X=1$. As a result, by taking only the lowest order term brought by $H_\mathrm{per}(t)$ into account, we arrive at the third-order van Vleck effective Hamiltonian in the perturbed case as follows:
\begin{eqnarray}
\tilde{D}_3 &\simeq& \sum_m \left( a_\mathrm{o} \tau_m^x +b_\mathrm{e} \tau_m^z \tau_{m+1}^z + h_z \tau_m^z \right) \nonumber \\
&\qquad& +  \sum_m \left( a_\mathrm{e} \tilde{\tau}_m^x +b_\mathrm{o} \tilde{\tau}_m^z \tilde{\tau}_{m+1}^z + h_z \tilde{\tau}_m^z \right).
\end{eqnarray}
Therefore, if the corresponding Kitaev chain composed of odd (even) sites is topologically nontrivial, the perturbation breaks the ground-state degeneracy of the corresponding transverse Ising chain described by $\tau_m^z$ ($\tilde{\tau}_m^z$), which results in the finite order parameter $\bra{\mathrm{GS}} \tau_m^z \ket{\mathrm{GS}}$ ($\bra{\mathrm{GS}} \tilde{\tau}_m^z \ket{\mathrm{GS}}$) where $\ket{\mathrm{GS}}$ is the ground state under $\tilde{H}_\mathrm{eff}$.

Now, let us describe how each of the four distinct phases is characterized by the real-time dynamics. First, we prepare the initial state by the ground state of the effective Hamiltonian $\tilde{H}_\mathrm{eff}$. This means that spontaneous symmetry breaking is assumed to take place for the system hosting nontrivial phases. Such symmetry-broken states can be introduced by an infinitesimal perturbation which breaks both the $\mathbb{Z}_2$ symmetry $X_\mathrm{all}$ and the discrete time translation symmetry such as $H_\mathrm{per}(t)$. They can also be attributed to measurement or decoherence. After preparing the symmetry-broken initial state, the system evolves under the $T$-periodic Hamiltonian $H(t)$, and then we focus on the stroboscopic dynamics of the order parameter
\begin{equation}
\left< \tau (nT) \right> \equiv \frac{1}{L} \sum_{m}^{L/2} \bra{\mathrm{GS}} U^\dagger (T)^n (\tau_m^z +\tilde{\tau}_m^z) U(T)^n \ket{\mathrm{GS}}.
\end{equation}
Since $\ket{\mathrm{GS}}$ is an eigenvector of $U(2T)$, we obtain
\begin{equation}
\left< \tau (2mT) \right> =  \frac{1}{L} \sum_{m}^{L/2} \bra{\mathrm{GS}} \tau_m^z \ket{\mathrm{GS}}
+  \frac{1}{L} \sum_{m}^{L/2} \bra{\mathrm{GS}} \tilde{\tau}_m^z \ket{\mathrm{GS}},
\end{equation}
for $m \in \mathbb{N}$. The first and the second term, which we denote $\phi_\mathrm{o}/2$ and $\phi_\mathrm{e}/2$ below, represent the order parameters of the corresponding transverse Ising chains which respectively characterize the corresponding Kitaev chains composed of odd and even sites. On the other hand, if the coarse-grained stroboscopic dynamics at $t=(2m+1)T$ is focused on, with the zero-th order approximation by
\begin{eqnarray}
U(T) &=& \mathcal{T} \exp \left\{ -i \int_0^{T} (H_0(t)+V(t))dt \right\} \nonumber \\
&\simeq&  \mathcal{T} \exp \left\{ -i \int_0^{T} H_0(t)dt \right\} = X_\mathrm{odd},
\end{eqnarray}
the following relation is obtained,
\begin{eqnarray}
\left< \tau ((2m+1)T) \right> &\simeq&  \frac{1}{L} \sum_{m}^{L/2} \bra{\mathrm{GS}} X_\mathrm{odd} \tau_m^z X_\mathrm{odd} \ket{\mathrm{GS}} \nonumber \\
&\quad& +  \frac{1}{L} \sum_{m}^{L/2} \bra{\mathrm{GS}} X_\mathrm{odd} \tilde{\tau}_m^z X_\mathrm{odd} \ket{\mathrm{GS}} \nonumber \\
&=& \frac{1}{L} \sum_{m}^{L/2} \left( - \bra{\mathrm{GS}} \tau_m^z \ket{\mathrm{GS}}
+ \bra{\mathrm{GS}} \tilde{\tau}_m^z \ket{\mathrm{GS}} \right) \nonumber \\
&=& \frac{1}{2}(- \phi_\mathrm{o} + \phi_\mathrm{e}).
\end{eqnarray}
Thus, $\left< \tau (nT) \right>$ shows a $2T$-periodic oscillation, in which its mean value and its amplitude are $\phi_\mathrm{e}/2$ and $\phi_\mathrm{o}/2$, respectively. As a result, whether the mean and the amplitude of $\left< \tau (nT) \right>$ are zero or nonzero tells us which phase the system lies in, from the following correspondence,
\begin{center}
$(Z_\mathrm{o},Z_\mathrm{e}) = (0,0)$ $\Leftrightarrow$ $\phi_\mathrm{o}=0$, $\phi_\mathrm{e}=0$, \\
$(Z_\mathrm{o},Z_\mathrm{e}) = (1,0)$ $\Leftrightarrow$ $\phi_\mathrm{o} \neq 0$, $\phi_\mathrm{e}=0$, \\
$(Z_\mathrm{o},Z_\mathrm{e}) = (0,1)$ $\Leftrightarrow$ $\phi_\mathrm{o}=0$, $\phi_\mathrm{e} \neq 0$, \\
$(Z_\mathrm{o},Z_\mathrm{e}) = (1,1)$ $\Leftrightarrow$ $\phi_\mathrm{o} \neq 0$, $\phi_\mathrm{e} \neq 0$.
\end{center}

We also numerically calculate the real-time dynamics to confirm the above conditions. Here, since the observable $\sum_m (\tau_m^z+\tilde{\tau}_m^z)/L$ is nonlocal, we employ the observable $(\tau_1^z+\tilde{\tau}_2^z)/2=(\sigma_1^z+\sigma_1^x \sigma_2^z)/2$ instead with assuming that the effect of the boundary is small. Figure \ref{Fig_dynamics} shows the numerical results of the stroboscopic dynamics of the order parameter
\begin{equation}
A(nT) = \frac{1}{2} \bra{\mathrm{GS}} U^\dagger(T)^n (\sigma_1^z + \sigma_1^x \sigma_2^z) U(T)^n \ket{\mathrm{GS}}
\end{equation}
for each set of parameters $g, \varepsilon$ in the four different phases. In the case of $g=-\gamma$ and $\varepsilon=0$ in the nontrivial phase $(Z_\mathrm{o},Z_\mathrm{e})=(1,1)$ [See Fig. \ref{Fig_dynamics} (d)], the observable $A(nT)$ shows a $2T$-periodic oscillation with its nonzero mean, and hence satisfying the above statement. As well, the dynamics in the remaining three regimes [See Fig. \ref{Fig_dynamics} (a), (b), and (c)] reflect which phase the system lies in, and hence we can conclude that SPT phases of the system can be identified by tracking the order parameter $A(nT)$. 

\begin{figure}
\begin{center}
    \includegraphics[height=4cm, width=8.5cm]{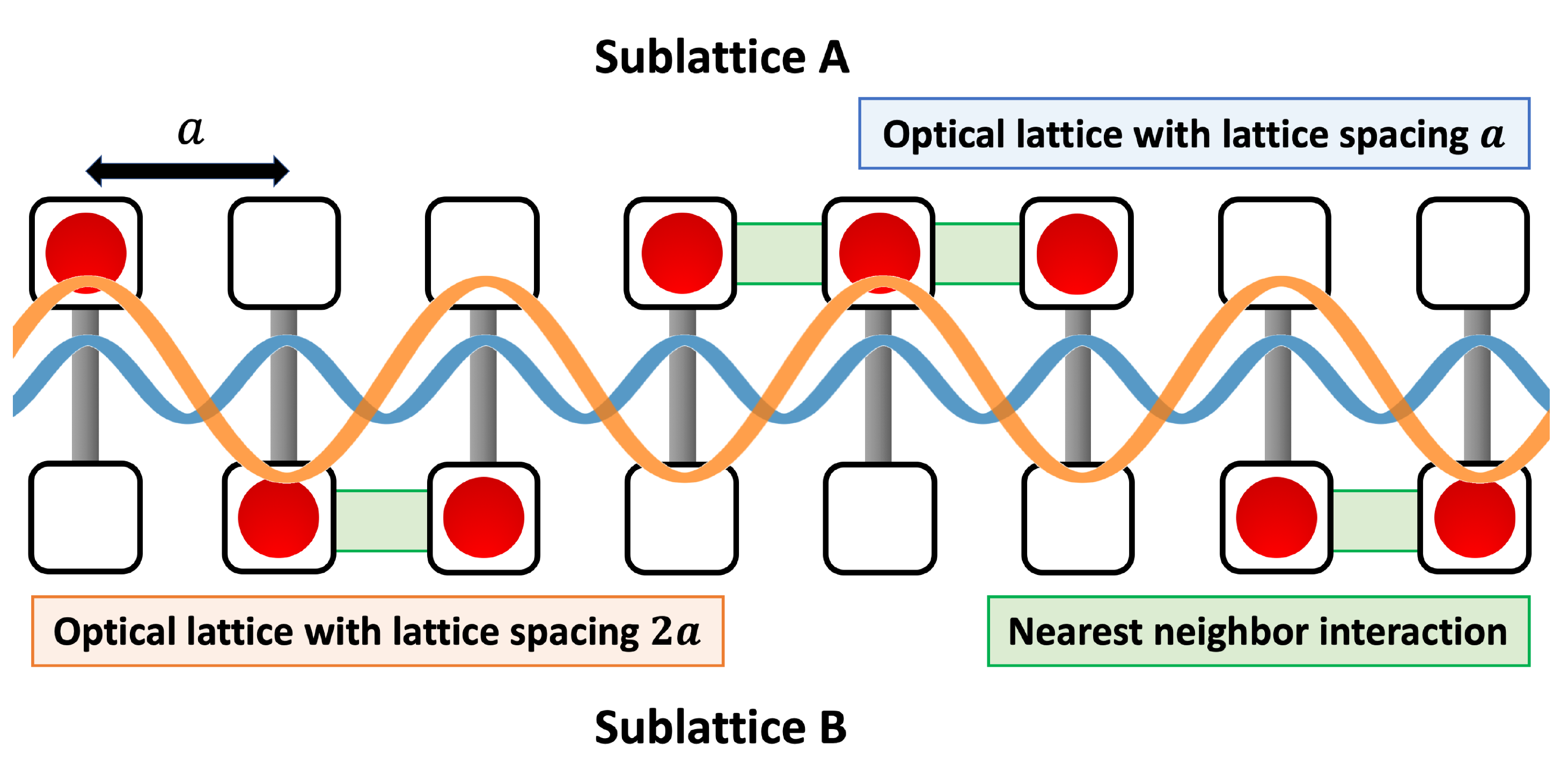}
    \caption{Possible experimental realization in ultracold atoms. The optical lattice with the latice spacing $a$ (blue wave) and the one with the lattice spacing $2a$ (orange wave) located between the two sublattices can reproduce the uniform fields $V_g+V_h(t)$ and the spatially-modulated $\pi$-pulse $H_0(t)$. The nearest neighbor interaction realizes the Ising interaction $V_\mathrm{int}(t)$.}
    \label{Fig_experiment}
  \end{center}
\end{figure}

We also would like to note about $2T$-periodic oscillations which appear in the nontrivial phases $Z_\mathrm{o}=1$. It should be noted that the observable oscillates with the period $2T$ in spite of the fact that the original Hamiltonian is $T$-periodic. Moreover, this $2T$-periodic oscillation is robust to perturbations which do not break the $\mathbb{Z}_2$ symmetry $X_\mathrm{all}$ and $T$-periodicity, since the oscillation is maintained as long as $Z_\mathrm{o}=1$ is satisfied. Thus, this behavior for $Z_\mathrm{o}=1$ is similar to the one in discrete time crystals (DTCs), which host spontaneous breaking of discrete time translation symmetry \cite{Khemani2016,Russomanno2017SPT}. Our model is different from DTCs in that this time-crystalline-like oscillation in our model only lasts in the prethermal regime before the thermalization to infinite temperature states, though DTCs show an everlasting period $n$-tupling in the thermodynamic limit. This behavior is similar to the one of prethermal discrete time crystals \cite{Else2017,Mizuta2019,Else2019}. Another difference is that the order parameter $\left< \tau(nT) \right>$ is nonlocal due to Jordan Wigner transformation unlike DTCs.

To summarize this section, we can identify each of the four distinct SPT phases from the stroboscopic dynamics of the nonlocal order parameter $\left< \tau (nT) \right>$ with $n \in \mathbb{Z}^{+}$. SPT phases discussed through this paper have been considered in terms of the coarse-grained stroboscopic dynamics at $t=2mT$ ($m \in \mathbb{Z}^+$), since they are defined by the effective Hamiltonian $H_\mathrm{eff}$, given by Eq. (\ref{H_eff}), which describes only the coarse-grained stroboscopic dynamics. Therefore, it should be noted that these SPT phases defined in the coarse-grained stroboscopic dynamics can be detected from the stroboscopic dynamics.


\section{Summary and Discussions}

In this paper, we propose a way to realize nontrivial topological phases protected by $\mathbb{Z}_2 \times \mathbb{Z}_2$ symmetry in a $\mathbb{Z}_2$ symmetric system by a combination of a resonant drive and high-frequency drives. The resonant drive is responsible for adding the emergent $\mathbb{Z}_2$ symmetry to the system, thereby making the system a platform for SPT phases under the $\mathbb{Z}_2 \times \mathbb{Z}_2$ symmetry as long as the system respects $\mathbb{Z}_2$. The high-frequency drives are responsible for generating topologically nontrivial terms via higher-order terms of the van Vleck expansion. The coexistence of such drives with the dual energy scales enables us to realize and control SPT phases under $\mathbb{Z}_2 \times \mathbb{Z}_2$ symmetry on the $\mathbb{Z}_2$ symmetric system in a robust and simple way by a static transverse field and a slight deviation of the resonant drive. We have also provided a way to detect the phases by a discrete-time-crystalline-like period-doubling of order parameters. In general, we can add a robust $\mathbb{Z}_N$ symmetry by a resonant drive and also treat higher-dimensional systems. Therefore, we expect that more complex SPT phases in higher dimensions, which are usually difficult to observe in equilibrium setups, could be realized by a combination of resonant and high-frequency drives in a way similar to our method in this paper. This is left for future work. 

Finally, we would like to comment on possible experimental platforms for realizing our model and performing the control of SPT phases. One of the key points is how to implement the resonant drive $H_0(t)$ given by Eq. (\ref{H_0_g}), since it should selectively drive only the odd sites. Such a spatially-modulated drive would be possible in atomic, molecular and optical (AMO) systems such as trapped ions thanks to their high controllability \cite{Zhang2017}. Here, we propose a concrete setup using cold atoms as the other promising platform for our proposal. We consider a one-dimensional ladder lattice composed of two sublattices $A$ and $B$ (Fig. \ref{Fig_experiment}). The initial state is prepared so that only one of the two sublattices is occupied by a hard-core boson at each site $j$. Then, each site $j$ has a spin degree of freedom with $S=1/2$, characterized by which sublattice is occupied. With this setup, Ising interaction corresponds to interactions between particles in nearest neighbor sites, which can be controlled by Feshbach resonance. On the other hand, hopping between the two sublattices effectively play a role of transverse field, which can be induced by adding an optical lattice between the two sublattices. The transverse field acting only on the even sites can be realized by using an optical lattice which has twice the lattice spacing of the ladder.

\begin{acknowledgments}
This work is supported by a Grant-in-Aid for Scientific
Research on Innovative Areas ``Topological Materials Science''
(KAKENHI Grant No. JP15H05855) and also JSPS KAKENHI
(Grants No. JP16J05078, No. JP18H01140, and JP19H01838). K. M. is supported by Doctoral Program for World-leading Innovative \& Smart Education,
Ministry of Education, Culture, Sports, Science and Technology.
K. T. thanks JSPS for support from Overseas Research Fellowship. \end{acknowledgments}


\begin{thebibliography}{44}
\bibitem [{Nof()}]{NofSite}%
  \BibitemOpen
  \href@noop {} {}\bibinfo {note} {In order to satisfy the condition Eq.
  (\ref{assumption2}) with $N=2$, the number of the odd sites should be even.
  For simplicity, we assume that the numebr of the sites is a multiple of
  four.}\BibitemShut {Stop}%
\end{thebibliography}

\providecommand{\noopsort}[1]{}\providecommand{\singleletter}[1]{#1}%

\clearpage

\newcommand{\wsq}{\qquad $\square$}
\newtheorem*{proof*}{Proof}
\newtheorem{lemma}{Lemma}
\renewcommand{\thesection}{A\arabic{section}}
\renewcommand{\theequation}{A\arabic{equation}}
\setcounter{equation}{0}
\renewcommand{\thefigure}{A\arabic{figure}}
\setcounter{figure}{0}
\setcounter{section}{0}

\onecolumngrid
\begin{center}
 {\large 
 {\bfseries Appendix }}
\end{center}
\vspace{10pt}

\section*{The van Vleck effective Hamiltonian of the model}
In this section, we show the derivation of the van Vleck effective Hamiltonian $D_3$ from the original Hamiltonian $H(t)$ given by Eq. (\ref{Hamiltonian}). First of all, the Fourier components of $V(t)=V_\mathrm{int}(t)+V_h(t)+V_g$, denoted by $\{ V_m \}$, can be calculated based on Eqs (\ref{Fourier}) and (\ref{U_0}), as follows,
\begin{eqnarray}
V_0 &=& g \sum_{j=1}^L \sigma_j^x, \label{A1} \\
V_{2} &=& V_{-2} = \frac{h}{2} \sum_{j=1}^L \sigma_j^x, \label{A2} \\
V_m &=& J \sum_{j=1}^{L-1} \sigma_j^z \sigma_{j+1}^z \frac{e^{im\omega\tau/2}-1}{im\pi} \quad (m:\mathrm{odd}), \label{A3} \\
V_m &=& 0 \quad (\mathrm{otherwise}). \label{A4}
\end{eqnarray} 
Since this model satisfies $[V_{-m},V_{m}]=0$, the second order term $D_\mathrm{vV}^{(2)}$ given by Eq. (\ref{vanVleck2}) dissappears. Then the third order effective Hamitonian $D_3$ is
\begin{equation} \label{A5}
D_3 = D_\mathrm{vV}^{(1)} + D_\mathrm{vV}^{(3)}
= g \sum_{j=1}^L \sigma_j^x + 4 \sum_{m\neq0}\sum_{n\neq0,m} \frac{[[V_{-m},V_{m-n}],V_n]}{3mn\omega^2}.
\end{equation}
In the second equality, we neglect the first term of $D_\mathrm{vV}^{(3)}$ since the third order term which includes $V_0$ is small compared to the second term due to the assumption Eq. (\ref{g_small}). From now on, we calculate the second term in Eq. (\ref{A5}), which we denote $D$ below. Focusing on the indices of the three Fourier components $V_{-m}$, $V_{m-n}$, and $V_{n}$, the summation of them gives zero. Thus, due to Eqs. (\ref{A1}) $\sim$ (\ref{A4}), the terms which satisfy $m-n=\pm2$ and $m:\mathrm{odd}$ have nonzero contributions. This results in
\begin{eqnarray}
D &=& \frac{4}{3\omega^2} \sum_{m:\mathrm{odd}} \frac{[[V_{-m},V_2],V_{m-2}]}{m(m-2)} +  \frac{4}{3\omega^2} \sum_{m:\mathrm{odd}} \frac{[[V_{-m},V_{-2}],V_{m+2}]}{m(m+2)} \nonumber \\
&=& \frac{2J^2 h}{3\omega^2} C \left[ \left[ \sum_{j=1}^{L-1} \sigma_j^z \sigma_{j+1}^z,\sum_{j=1}^L \sigma_j^x \right], \sum_{j=1}^{L-1} \sigma_j^z \sigma_{j+1}^z \right],
\end{eqnarray}
where the real constant $C$ is given by the following equation,
\begin{equation} \label{A7}
C = \sum_{m:\mathrm{odd}}  \frac{(e^{-im\omega\tau/2}-1)(e^{i(m-2)\omega\tau/2}-1)}{m^2(m-2)^2\pi^2} + \sum_{m:\mathrm{odd}}  \frac{(e^{-im\omega\tau/2}-1)(e^{i(m+2)\omega\tau/2}-1)}{m^2(m+2)^2\pi^2}.
\end{equation}
The commutator in $D$ is calculated as follows,
\begin{eqnarray}
\left[ \sum_{j=1}^{L-1} \sigma_j^z \sigma_{j+1}^z,\sum_{j=1}^L \sigma_j^x \right] &=& \sum_{j=1}^{L-1} \left( [\sigma_j^z,\sigma_j^x]\sigma_{j+1}^z +  \sigma_j^z[\sigma_{j+1}^z,\sigma_{j+1}^x] \right) \nonumber \\
&=& 2i \sum_{j=1}^{L-1} ( \sigma_{j}^y \sigma_{j+1}^z + \sigma_j^z \sigma_{j+1}^y ),
\end{eqnarray}
and
\begin{eqnarray}
\left[ \left[ \sum_{j=1}^{L-1} \sigma_j^z \sigma_{j+1}^z,\sum_{j=1}^L \sigma_j^x \right], \sum_{j=1}^{L-1} \sigma_j^z \sigma_{j+1}^z \right] &=&
2i \sum_{j=1}^{L-1} [\sigma_j^y,\sigma_j^z] (\sigma_{j+1}^z)^2 + 4i \sum_{j=2}^{L-1} \sigma_{j-1}^z [\sigma_j^y,\sigma_j^z] \sigma_{j+1}^z + 2i \sum_{j=1}^{L-1} (\sigma_j^z)^2 [\sigma_{j+1}^y,\sigma_{j+1}^z] \nonumber \\
&=& -8 \left( \sum_{j=1}^L \sigma_j^x + \sum_{j=2}^{L-1} \sigma_{j-1}^z \sigma_{j}^x \sigma_{j+1}^z \right) + 4(\sigma_1^x +\sigma_L^x).
\end{eqnarray}
The last term $4(\sigma_1^x +\sigma_L^x)$, which locally acts on the edge $j=1,L$, is neglected in the main text, since such a term becomes irrelevant for topological phases when the system size $L$ is large.

Next, we calculate the real constant $C$, which is given by Eq. (\ref{A7}),
\begin{eqnarray}
C &=& \sum_{l \in \mathbb{Z}} \frac{(e^{-il\omega\tau-i\omega\tau/2}-1)(e^{il\omega\tau-i\omega\tau/2}-1)}{(2l-1)^2(2l+1)^2\pi^2} + \sum_{l \in \mathbb{Z}} \frac{(e^{-il\omega\tau+i\omega\tau/2}-1)(e^{il\omega\tau+i\omega\tau/2}-1)}{(2l-1)^2(2l+1)^2\pi^2} \nonumber \\
&=& \frac{4}{\pi^2} \cos \frac{\omega\tau}{2} \sum_{l \in \mathbb{Z}} \frac{\cos (\omega\tau/2) - \cos l\omega\tau}{(2l-1)^2(2l+1)^2}
\end{eqnarray}
By using the formula
\begin{equation}
\sum_{l=1}^\infty \frac{1}{(2l-1)^2(2l+1)^2}=\frac{\pi^2-8}{16},
\end{equation}
we obtain
\begin{eqnarray}
C &=& \frac{1}{2} \cos^2 \frac{\omega\tau}{2} - \frac{4}{\pi^2} \cos \frac{\omega\tau}{2} - \frac{8}{\pi^2} \cos \frac{\omega\tau}{2} \sum_{l=1}^\infty  \frac{\cos l\omega\tau}{(2l-1)^2(2l+1)^2}. \label{A12}
\end{eqnarray}
Here, we calculate the second summation as follows,
\begin{eqnarray}
\sum_{l=1}^\infty  \frac{\cos l\omega\tau}{(2l-1)^2(2l+1)^2} &=&
- \frac{1}{2} \sum_{l=1}^\infty \frac{\cos l\omega\tau}{(2l-1)(2l+1)} + \frac{1}{4} \sum_{l=1}^\infty \frac{\cos l\omega\tau}{(2l-1)^2} + \frac{1}{4} \sum_{l=1}^\infty \frac{\cos l\omega\tau}{(2l+1)^2} \nonumber \\
&=& \frac{\pi}{8} \sin \frac{\omega\tau}{2} -\frac{1}{4} + \frac{1}{4} \sum_{l=1}^\infty \frac{\cos l\omega\tau}{(2l-1)^2} + \frac{1}{4} \sum_{l=1}^\infty \frac{\cos l\omega\tau}{(2l+1)^2}.
\end{eqnarray}
In the second equality, we have used the formula
\begin{equation}
\sum_{l=1}^\infty \frac{\cos 2lx}{(2l-1)(2l+1)} = \frac{1}{2} - \frac{\pi}{4} \sin x
\end{equation}
for $0\leq x\leq \pi$. With using the relation
\begin{equation}
\sum_{l=1}^\infty \frac{\cos (2l-1) x}{(2l-1)^2} = \frac{\pi}{4} \left( \frac{\pi}{2} -x \right)
\end{equation}
for $0 \leq x \leq \pi$, the remaining terms are obtained:
\begin{eqnarray}
\frac{1}{4} \sum_{l=1}^\infty \frac{\cos l\omega\tau}{(2l-1)^2} + \frac{1}{4} \sum_{l=1}^\infty \frac{\cos l\omega\tau}{(2l+1)^2}
&=& \frac{1}{4} \sum_{l=1}^\infty \frac{\cos l\omega\tau + \cos\{(l-1)\omega\tau\}}{(2l-1)^2} - \frac{1}{4} \nonumber \\
&=& \frac{1}{2} \cos \frac{\omega\tau}{2} \sum_{l=1}^\infty \frac{\cos \{(2l-1)\omega\tau/2\}}{(2l-1)^2} -\frac{1}{4} \\
&=& \frac{\pi}{16} (\pi-\omega\tau) \cos \frac{\omega\tau}{2} - \frac{1}{4},
\end{eqnarray}
and 
\begin{eqnarray}
\sum_{l=1}^\infty  \frac{\cos l\omega\tau}{(2l-1)^2(2l+1)^2} &=& \frac{\pi}{8} \sin \frac{\omega\tau}{2} +  \frac{\pi}{16} (\pi-\omega\tau) \cos \frac{\omega\tau}{2} - \frac{1}{2}.
\end{eqnarray}
Substituting this into Eq. (\ref{A12}) results in the real constant $C$ given by
\begin{eqnarray}
C &=& \frac{1}{4\pi} \{ \omega\tau (1+\cos \omega \tau) - 2 \sin \omega \tau \}.
\end{eqnarray}
Finally, we arrive at the third order van Vleck effective Hamiltonian $D_3$:
\begin{eqnarray}
D_3 &=& (g+\gamma) \sum_{j=1}^L \sigma_j^x + \gamma \sum_{j=2}^{L-1} \sigma_{j-1}^z \sigma_j^x \sigma_{j+1}^z -\frac{\gamma}{2}(\sigma_1^x+\sigma_L^x), \\
\gamma &=& \frac{4J^2h}{3\pi\omega^2} \{ 2\sin \omega \tau - \omega\tau (1+\cos \omega \tau) \},
\end{eqnarray}
which we have provided as the effective Hamiltonian Eq. (\ref{D_3}) in the main text.

\end{document}